\documentclass{pasj00}

\begin{document}
\SetRunningHead{Kubota et al. 2006}{Absorption lines in 4U~$1630-472$}
\Received{2006/07/25}
\Accepted{2006/10/12}

\title{Suzaku discovery of iron absorption lines in 
outburst spectra of the X-ray transient 4U~$1630-472$
 }



\author{
Aya \textsc{Kubota}\altaffilmark{1},
Tadayasu \textsc{Dotani}\altaffilmark{2}, 
Jean \textsc{Cottam}\altaffilmark{3}, 
Taro \textsc{Kotani}\altaffilmark{4}, \\
Chris \textsc{Done}\altaffilmark{2,5},
Yoshihiro \textsc{Ueda}\altaffilmark{6}, 
Andrew C. \textsc{Fabian}\altaffilmark{7},
Tomonori \textsc{Yasuda}\altaffilmark{8},\\ 
Hiromitsu \textsc{Takahashi}\altaffilmark{8}, 
Yasushi \textsc{Fukazawa}\altaffilmark{8}, 
Kazutaka \textsc{Yamaoka}\altaffilmark{9},\\
Kazuo \textsc{Makishima}\altaffilmark{1,10},
Shinya \textsc{Yamada}\altaffilmark{10},
Takayoshi \textsc{Kohmura}\altaffilmark{11},
and Lorella \textsc{Angelini}\altaffilmark{3}
}

\altaffiltext{1}{
Institute of Physical and Chemical Research (RIKEN), 
2-1 Hirosawa, Wako, Saitama 351-0198 
}
\email{aya@crab.riken.jp}
\altaffiltext{2}{
Institute of Space and Astronautical Science, 
Japan Aerospace Exploration Agency, \\
3-1-1 Yoshinodai, Sagamihara, Kanagawa 229-8510
}
\altaffiltext{3}{
Exploration of the Universe Division,\\
NASA Goddard Space Flight Center, Greenbelt, MD 20771, USA
}
\altaffiltext{4}{
Department of Physics, Tokyo Tech, 2-12-1 O-okayama, Meguro, Tokyo 152-8551
}
\altaffiltext{5}{
Department of Physics, University of Durham, 
South Road, Durham, DH1 3LE, UK
}
\altaffiltext{6}{
Department of Astronomy, Kyoto University, Sakyo-ku, Kyoto 606-8502
}
\altaffiltext{7}{
Institute of Astronomy, Madingley Road, Cambridge CB3 0HA, UK}
\altaffiltext{8}{
Department of Physics, Hiroshima University,\\
1-3-1 Kagamiyama, Higashi-Hiroshima, Hiroshima 739-8526
}
\altaffiltext{9}{
Department of Physics, Aoyama Gakuin University, Sagamihara, Kanagawa 229-8558
}
\altaffiltext{10}{
Department of Physics, University of Tokyo, 
7-3-1 Hongo, Bunkyo-ku, Tokyo 113-0033
}
\altaffiltext{11}{Physics Department, Kogakuin University 
2665-1, Nakano-cho, Hachioji, Tokyo, 192-0015 }

%


\maketitle

\begin{abstract}
We present the results of six Suzaku observations of the recurrent
black hole transient 4U~$1630-472$ during its decline from outburst
from February 8 to March 23 in 2006. All observations show the typical
high/soft state spectral shape in the 2--50~keV band, roughly described
by an optically thick disk spectrum in the soft energy band plus a weak
power-law tail that becomes dominant only above $\sim$20~keV.  The disk
temperature decreases from 1.4~keV to 1.2~keV as the flux decreases by a
factor 2, consistent with a constant radius as expected for
disk-dominated spectra. All the observations reveal significant
absorption lines from highly ionized (H-like and He-like) iron K$\alpha$
at 7.0~keV and 6.7~keV.  The brightest datasets also show significant
but weaker absorption structures between 7.8~keV and 8.2~keV, which we identify as a
blend of iron K$\beta$ and nickel K$\alpha$ absorption lines. 
The energies of these absorption lines suggest a blue shift with 
an outflow velocity of $\sim1000~{\rm km~s^{-1}}$.
The H--like iron K$\alpha$ equivalent width remains approximately constant
at $\sim30$~eV over all the observations, while that of the He--like K$\alpha$
line increases from 7~eV to 20~eV. Thus the ionization state of the material
decreases, as expected from the decline in flux. We fit the profiles
with Voigt functions (curve of growth) to derive absorbing columns, then use
these together with detailed photo-ionization calculations to derive
physical parameters of the absorbing material.
The data then constrain the velocity dispersion of the absorber to 200--$2000~{\rm
km~s^{-1}}$, 
and the size of the plasma as $\sim10^{10}~{\rm cm}$ 
assuming a source distance of 10~kpc.


\end{abstract}

\section{Introduction}
In recent years, a growing number of X-ray binaries have been found to exhibit absorption lines from highly ionized elements (e.g., \cite{boirin04} and references therein; \cite{church05}).  These systems range from microquasars such as GRO~J1655$-$40 \citep{ueda98,yamaoka01,miller06} and GRS~1915+105 \citep{kotani00,lee02} to low-mass X-ray binaries such as GX~13+1 \citep{ueda01,sidoli02}, X~1658$-$298 \citep{sidoli01}
and X~1254$-$690 \citep{boirin03}.
These systems are viewed at high inclination angles, and 
the absorption structure is visible throughout the orbital period (e.g. \cite{yamaoka01,sidoli01}; 2002).  
The absorption features are therefore
thought to originate in material that is associated with and extends above the outer accretion
disk. This is illuminated by the X-rays produced from the innermost regions of the accretion
flow (both the disk and hard X-ray coronal emission). The reprocessed emission and scattered
flux from the extended wind can be seen directly in the Accretion Disk Corona sources, where
the intrinsic X-rays are obscured (e.g. Kallman et al. 2003), but for the majority of highly
inclined sources, the wind material is seen in absorption against the much brighter intrinsic
central X-ray source. 
Multiple absorption lines seen there then give an excellent
probe of the physical conditions in the wind (e.g. Ueda et al. 2004; Miller et al. 2006),
 where the spectra
indicate the presence of significant amounts of highly ionized outflowing 
material.
Such accretion disk winds can be produced by thermal driving (e.g., Begelman et al. 1983), 
and/or by magnetic forces \citep{blandford82}
as inferred by \citet{miller06} from Chandra HETGS data of GRO~J$1655-40$.

4U~1630$-$472 is a black hole candidate (e.g., \cite{tanaka95, mc03}), known for X-ray outbursts that repeat with an interval of roughly 600--690 days (Jones et al. 1976; Parmar, Angelini \& White 1995).  
The source has been observed with every major mission from 
EXOSAT (\cite{kuulkers97}; Parmar et al. 1986), Ginga (Parmar et al. 1997), BATSE CGRO
(Bloser et al. 1996), to  Beppo-SAX (Oosterbroek et al. 1998), 
RXTE (Tomsick et al.~1998; Kuulkers et al.~1998; Hjellming et al. 1999; 
Cui et al. 2000; Dieters et al.~2000; Trudolyubov et al. 2001; Kalemci et al. 2004; 
\cite{abe05}),  ASCA (\cite{abe05}), and recently  Chandra (J. Miller, private communication).  
These X-ray observations showed that 
the source is absorbed heavily with $N_{\rm H}=(5$--$12)\times10^{22}~{\rm cm^{-2}}$ 
(Tomsick et al. 1998; Kuulkers et al. 1997; Cui et al. 2000; Trudolyubov et al. 2001; \cite{abe05}). 
While the IR counterpart was found during a 1998 outburst by Augusteijn et al. (2001),
no optical counterpart is known for 4U~1630$-$472, most likely due to its high reddening and location in a crowded star field (Parmar, Stella \& White 1986).  
Thus, the distance, inclination angle, and mass of the compact object 
are unknown for 4U~1630$-$472.
This source is classified as a black hole not from dynamic mass measurements, but because of the similarity of its X-ray spectral and timing properties to those of systems with measured black hole masses.

This source was observed six times with Suzaku during its most recent outburst as part of a program to study discrete spectral structures 
as a function of the changing accretion conditions.
The Suzaku monitoring observation led to successful detection of absorption lines 
on all occasions.
Though one of the absorption lines 
may have been detected by Chandra (J. Miller, private communication),
the Suzaku XIS data reveal the line with a much higher significance, thanks to its 
larger effective area at 7~keV than that of Chandra.
In this paper, we
focus mainly on the absorption line structures, and how these discrete features
evolve over the two months span of our observations.
Considering the large absorption column density, we assume the source distance, $D$, 
at $10$~kpc in this paper.
In addition, by comparison to the other black hole binaries reported to show absorption lines, 
an inclination angle, $i$, of $70^\circ$ is assumed.
Errors quoted in this paper represent 90\% confidence limit for 
a single parameter unless otherwise specified.

\section{Observations and data reduction}

\subsection{Observations}
An outburst of 4U~$1630-472$ was reported by the
RXTE All Sky Monitor (ASM; \cite{levine96}) on 2005 December.  From 2006 February 8
through March 23, we observed the source six times with a series of
21--23~ks exposures as part of the initial performance verification of 
Suzaku.  Suzaku is the 5th X-ray astronomy satellite of Japan launched
on July 10, 2005, with an M-V rocket from the Uchinoura space center
(Mitsuda et al.\ 2006).  
It carries 4 sets of X-ray telescopes (\cite{xrt06}) each with a focal-plane X-ray CCD
camera (XIS, X-ray Imaging Spectrometer; \cite{koyama06}) 
operating in the energy range of 0.2--12~keV, 
together with a non-imaging 
Hard X-ray Detector (HXD, Takahashi et al.\ 2006), which covers the 10--600 keV energy band with
Si PIN photo-diodes and GSO scintillation counters. 
Three of the XIS (XIS0, 2, 3) have front-illuminated (FI) CCDs, while
XIS1 utilizes a back-illuminated (BI) CCD,
achieving an improved soft X-ray
response but poorer hard X-ray sensitivity. 
Considering the  large column density to 4U~$1630-472$, we concentrate on the harder X-ray band in this paper, and only analyze the 
data from the FI CCDs. 

\begin{figure}
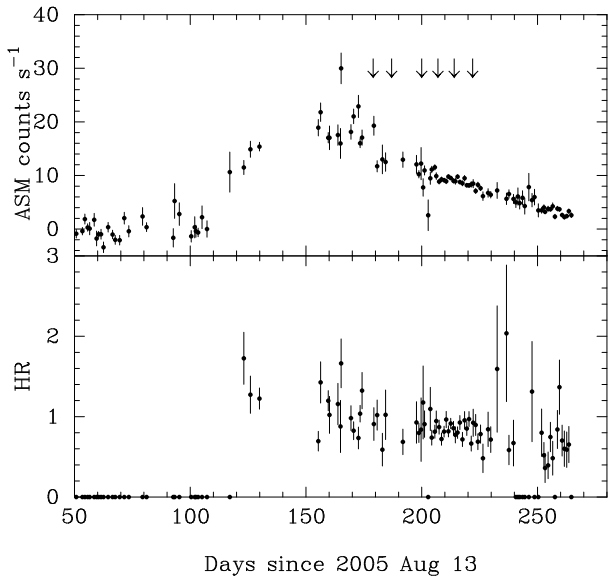

  \begin{center}
    \FigureFile(80mm,60mm){figure1.ps}
  \end{center}
  \caption{The 1.5--12~keV light curve of 4U~$1630-472$ (top panel), and hardness ratio of 5--12~keV to 3--5~keV  (bottom panel),obtained with the  RXTE ASM since 2005 August 13 (MJD 53595).
  The  Suzaku observations are indicated with down-arrows. }
  \label{fig:asm}
\end{figure}

Figure~\ref{fig:asm} shows a lightcurve of 4U~$1630-472$ obtained by the
 RXTE ASM.  The Suzaku observations are indicated with downward
arrows.  Because the optical axes of the XIS and the HXD are slightly offset,
the source was put at the center of the HXD field of view during the
observations, to maximize the HXD count rate, and to reduce the XIS count
rate as pileup is an issue for such a bright source (see below). The HXD was
operated in the nominal mode throughout the observations.
Because the source was very bright, the XIS
was operated with the 1/4 window option, in which a smaller field
of view of $17^\prime.8\times 4^\prime.5$ is read out every 2~s.
A burst option was added in the first four 
observations.  This decreases the effective exposure by a factor of two.
The editing modes of the FI CCDs (XIS023) were set to $3\times3$ and
$2\times2$ for high and medium data rates, respectively.
Table~\ref{tab:obs} summarizes the details of the observations.

\begin{table*}
  \caption{Suzaku Observations of 4U~$1630-472$}
  \label{tab:obs}
  \begin{center}
    \begin{tabular}{clllcrcc}
    \hline
    Epoch & \multicolumn{1}{c}{Start} & \multicolumn{1}{c}{End} & \multicolumn{1}{c}{XIS options} & \multicolumn{2}{c}{Exposure [ks]  $^*$}&
     \multicolumn{2}{c}{Flux~$[{\rm erg~s^{-1}~cm^{-2}}]$}\\ 
        	&	&	&	&total$^\dagger$ & High$^\dagger$ 	&2--10~keV	&20--40~keV\\ \hline
1	&Feb 8 8:14 	&Feb 9 3:20 	&1/4 window, 1-s burst	& 22.2&22.2&$4.6\times10^{-9}$&$6.9\times10^{-11}$\\
2	&Feb 15 23:08	&Feb 16 13:08	&1/4 window, 1-s burst	& 21.5&9.6&$4.0\times10^{-9}$&$6.7\times10^{-11}$\\

3	&Feb 28 23:00 	&Mar 1 16:43	&1/4 window, 1-s burst	& 21.3&21.3&$3.1\times10^{-9}$&$5.0\times10^{-11}$\\

4	&Mar 8 01:39	&Mar 8 17:31	&1/4 window, 1-s burst	& 21.2&20.7&$2.8\times10^{-9}$&$3.5\times10^{-11}$\\

5	&Mar 15 5:19	& Mar 15 17:51	&1/4 window	& 23.2&5.2&$2.5\times10^{-9}$&$2.6\times10^{-11}$\\

6	&Mar 23 9:57	&Mar 23 22:21	&1/4 window	& 21.7&15.4&$2.2\times10^{-9}$&$1.5\times10^{-11}$\\

\hline
   \multicolumn{7}{@{}l@{}}{\hbox to 0pt{\parbox{160mm}{\footnotesize
      \par\noindent
      \footnotemark[$*$] Without correction for the burst option. 
With the burst option, the net exposure is a half of the nominal exposure listed in this table.
      \par\noindent
      \footnotemark[$\dagger$] Net exposures of total (medium and high data rates) and only high data rate are shown. For FI-CCDs, observations were performed with $3\times3$ and $2\times2$ mode in data-rate High and 
      Med, respectively.
                }\hss}}
\end{tabular}
\end{center}
\end{table*}

\subsection{Data reduction}

For both the XIS and the HXD, we use version 0.7 screened data
provided by the  Suzaku team.
The screening of the version 0.7 XIS data are 
based on the following standard criteria: 
a) Only GRADE0, 2, 3, 4, 6 events are accumulated,
b) The time interval after passage through the South Atlantic Anomaly is 
larger than 436 seconds, 
c) The object is at least $5^\circ$ and $20^\circ$ 
above the rim of the Earth during night and day, respectively.
We limit our analysis to the $3\times3$ mode data (high data rate), 
because the $2\times2$ mode data
(medium data rate) are not well calibrated at this time.
During the 2nd and 5th observations, more than half of the exposure was acquired in $2\times2$ mode,  
and thus these data have poorer statistics.
The exposure times for the total and high data rate are summarized in Table~\ref{tab:obs}.

The XIS events were extracted from
a circular region with a radius of $4^\prime\!.3$ centered on the image peak.
This extraction circle is larger than the window size.
The effective extraction region is therefore the intersection of the window
and this circle.
The background was entirely
negligible throughout the six observations
(less than 1\% of the signal at 9~keV), 
so was not subtracted from the data.  Using XIS2 as a
reference, we find that the 2--10~keV count rate declined over the six
observations, with 
$124$, $112$, $90$,
$82$, $74$, and $64~{\rm cts~s^{-1}}$ respectively,
roughly corresponding to 4.6, 4.0, 3.1, 2.8, 2.5, and
$2.2\times10^{-9}~{\rm erg~s^{-1}~cm^{-2}}$ in the same energy band.  Within each observation, we find no significant intensity
variation intrinsic to the source.

The source was so bright during these observations that the XIS suffered
from photon pileup at the image center. 
A rough threshold above which the photon pileup becomes significant is 
$\sim\!100$~cts~exposure$^{-1}$ for a point source.
The count rates of 4U~$1630-472$ during the observations were around this level.
Furthermore, because 4U~$1630-472$ has a very soft energy spectrum, the effect
of photon pileup can be significant in the hard band spectra.
Therefore, we excluded a circular region of $30^{\prime\prime}$ at the
image center from the event extraction region to minimize these effects.
This did not remove the photon pileup completely,
and an artificial structure is seen as a hard excess above $\sim8$~keV
at the $<$5--10\% level.
Because photon pileup tends to produce X-ray events which have
twice (or a little more) as high an energy as the most abundant events 
(in this case, about 4 keV), we modeled the artificial structure as 
a gaussian centered at 10~keV with a fixed width of $\sigma = 1$~keV.
As seen in the next section, this phenomenological model works
quite well to compensate for the effect of pileup.

Because the event extraction region has a complex shape
(an annulus truncated with the window size), we calculated 
the effective area for each XIS sensor by using {\sc xissimarfgen} 
(version 2006-04-24), which calculates the arf file through 
Monte Carlo simulations.
The three FI CCDs have almost the same performance, 
so we have added their data to increase the statistics in the spectra.
Hereafter we refer to the summed data as XIS023.  We calculated 
the corresponding instrumental responses by summing 
the latest redistribution matrices of ae\_xi[0,2,3]\_20060213.rmf using the 
{\sc ftool} {\sc addrmf} and the {\sc arf} files using the {\sc ftool} {\sc addarf}. 

Though the main objective of this paper is the absorption structure in the XIS energy 
band, 
it is also useful to understand the continuum emission since this provides 
a knowledge of spectral states and photoionising flux.
We thus analyzed the HXD data using only the PIN data, 
since the source is dim in the GSO band.
Both high and medium data rate are used in the PIN data analysis.
The version 0.7 PIN data are screened by  
the following selection criteria: 
a) The object is at least $5^\circ$ above the earth rim,
b) The time interval after passage through the South 
Atlantic Anomaly is longer than 500 seconds,
c) The cutoff rigidity is greater than $8~{\rm GeV~c^{-1}}$.
The PIN background events were provided by the HXD team for each 
observation based on two different models, A and B (Kokubun et al.\ 2006).
Here we constructed the PIN background spectra based on model B
filtered by the same good time intervals.
The latest version of the response matrix, ae\_hxd\_pinhxnom\_20060419.rmf, 
was used in our spectral analysis.
Even if we allow 5\% systematic uncertainty in the background 
level, the source is clearly detected up to 40~keV and 35~keV in PIN data 
from the first four and last two observations, respectively.
Dead time fractions were calculated from the pseudo event rates (\cite{tt06,kokubun06})
as 5.3, 4.3, 4.6, 5.1, 4.6, and 4.6\%, in the first to last 
observations, respectively. 
After the dead time correction, the background subtracted PIN count 
rates are estimated to be 0.16, 0.13, 0.093, 0.065, 0.055, and 
0.043~${\rm cts~s^{-1}}$ in the range of 20--40~keV, corresponding to 
6.9, 6.7, 5.0, 3.5, 2.6, and $1.5\times10^{-11}~{\rm erg~s^{-1}~cm^{-2}}$. 

\subsection{Calibration uncertainties of the XIS data}

Because the XIS calibration is still in progress, we include a
brief description of the calibration uncertainties.
The XIS energy scale is relatively well calibrated with an uncertainty of less than 5~eV, or 0.2\% 
when the standard mode is used (\cite{koyama06}).
However, when a window option is applied, 
the gain can change slightly,
because the number of frame-store and parallel
transfers are different than in the standard mode.
This difference is not taken into account in the version 0.7 processing software.  
Since a window option was used
for all the observations of 4U~$1630-472$,
the XIS instrument team estimates that this produces a systematic gain difference
of $\sim$10~eV at 6~keV at the time of 4U~$1630-472$ observations.
When the window option is used, the onboard $^{55}$Fe calibration source is not 
directly visible but some events can be
scattered into the active window in XIS2, which has the strongest
calibration sources.   
The scattered $^{55}$Fe events are accumulated from the regions free from 
the source photons ($1/8$ of the window area for each side).
We obtain an average energy of 
$5891\pm20$~eV.
This is consistent with the $\sim$10~eV offset estimated above.  We also examined the data for a
possible gain shift due to the large surface brightness (local event density)
of the image, by comparing the absorption line energies in the
observed spectra for different radii of 
extraction annuli.  
We find no significant differences.   Thus 
a conservative estimate is that the XIS gain at the iron 
band is
correct to within 20~eV during the 4U~$1630-472$ observations.

The energy resolution of the XIS is gradually degrading due to the radiation damage on the CCD.
However, 
this effect is not yet included in the currently available response matrices.   
At the time of the 4U~$1630-472$ observations, a narrow
line at 6 keV would appear broadened with $\sigma \sim 40$~eV, 
where $\sigma$ is an equivalent gaussian width (standard deviation).
Thus, when we evaluate a
line width, we take this effect into account.
It is important to note that this does not affect the measurement of the
equivalent width.

The effective areas of the XIS and PIN are calibrated using observations of the Crab Nebula.
The accuracy is expected to be 5--10\%.  This is sufficient to estimate
various emission parameters of 4U~$1630-472$.
However, because of the very good statistics of the data, 
the systematic errors in calibration 
could degrade the model fitting significantly.
A major contribution to the systematic error is the calibration uncertainty at gold M-II edge in 
the instrumental response. To account for this residual, we added a gaussian at $\sim3.2$~keV 
with a fixed width of $\sigma=0.1$~keV.

\section{Analyses and results} 

\subsection{Broad band features}
\begin{figure*}
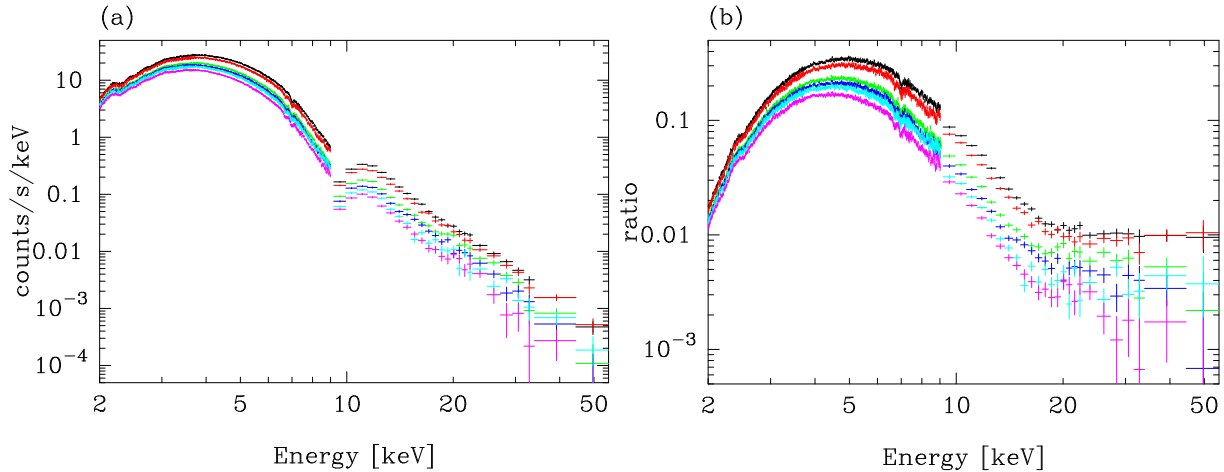

  \begin{center}
    \FigureFile(80mm,60mm){figure2a.ps}
      \FigureFile(80mm,60mm){figure2b.ps}
  \end{center}
  \caption{(a) The X-ray spectra of 4U~$1630-472$ obtained with the XIS023 and the HXD PIN. 
  The first to the last observations are 
shown with different colors as black, red, green, blue, right blue, and magenta.
(b) The ratio of 4U~$1630-472$ spectra to the Crab spectrum.}

  \label{fig:rawspec_xis-pin}
\end{figure*}

Figure~\ref{fig:rawspec_xis-pin}a shows the 2--50~keV spectra of 4U~$1630-472$ 
obtained with the XIS023  and the PIN.
We do not use data below 2~keV,
because of large interstellar absorption.
An instrument-independent way to 
view the spectra is to construct the ratio of the source spectra to the corresponding data
from the Crab Nebula, which has an approximately power-law spectrum with
$\Gamma=2.1$, and interstellar absorption of 
$N_{\rm H}=3\times10^{21}~{\rm cm^{-2}}$ (\cite{toor74, cani86}).  
Figure~\ref{fig:rawspec_xis-pin}b shows the Crab ratios calculated
using the Crab data obtained with Suzaku 
on 2006 April 5 at the HXD nominal position and the XIS operated with the 1/4 window option. 
This plot reveals several noticeable features in the broad band
source spectra.  
The source spectra are characterized by a dominant soft thermal component in
the XIS band, which is heavily absorbed, while the harder X-ray PIN
data clearly show a weak power-law tail.

The dominant soft component is generally observed from bright black hole binaries in the 
high/soft state \citep{tanaka95}, and is believed to be emission from the optically thick standard accretion disk 
(\cite{ss73}).
In the Crab ratio (figure~\ref{fig:rawspec_xis-pin}b), 
the weak power-law tail is roughly characterized as $\Gamma\sim2$ in the first two observations.
When the 20--50~keV PIN spectra of these two observations 
are fitted with the single power-law model 
under the current response matrix, 
the values of $\Gamma$ are obtained as $2.4\pm0.4$ and $2.5\pm0.5$, respectively.
The extrapolated 20--100~keV fluxes are estimated as $(1.4\pm0.3)\times 10^{-10}$ and 
$(1.1 \pm 0.3)\times 10^{-10}~{\rm erg~s^{-1}~cm^{-2}}$, respectively, which are 
only 3~\% of the absorbed disk flux in the range of 2--10~keV.
Though the steeper power-law tails may be suggested on 
the latter four observations from the Crab ratio plot 
(figure~\ref{fig:rawspec_xis-pin}b), 
they are difficult to quantify because of their faintness under the current background uncertainty.
Further detailed analysis of the HXD data will be presented in a separate paper.
 
The dominance of the disk and the weak power-law tail are both characteristic of the
high/soft state. 
In addition, the Crab ratios also reveal the presence of complex absorption structures
in the iron K band. 
In this paper, we
concentrate on these absorption line structures and their relation to
the accretion disk parameters. 
Hereafter, we perform detailed spectral analyses on the XIS data.

\subsection{XIS spectra with canonical multi-color disk and negative gaussian fits}

\begin{figure*}
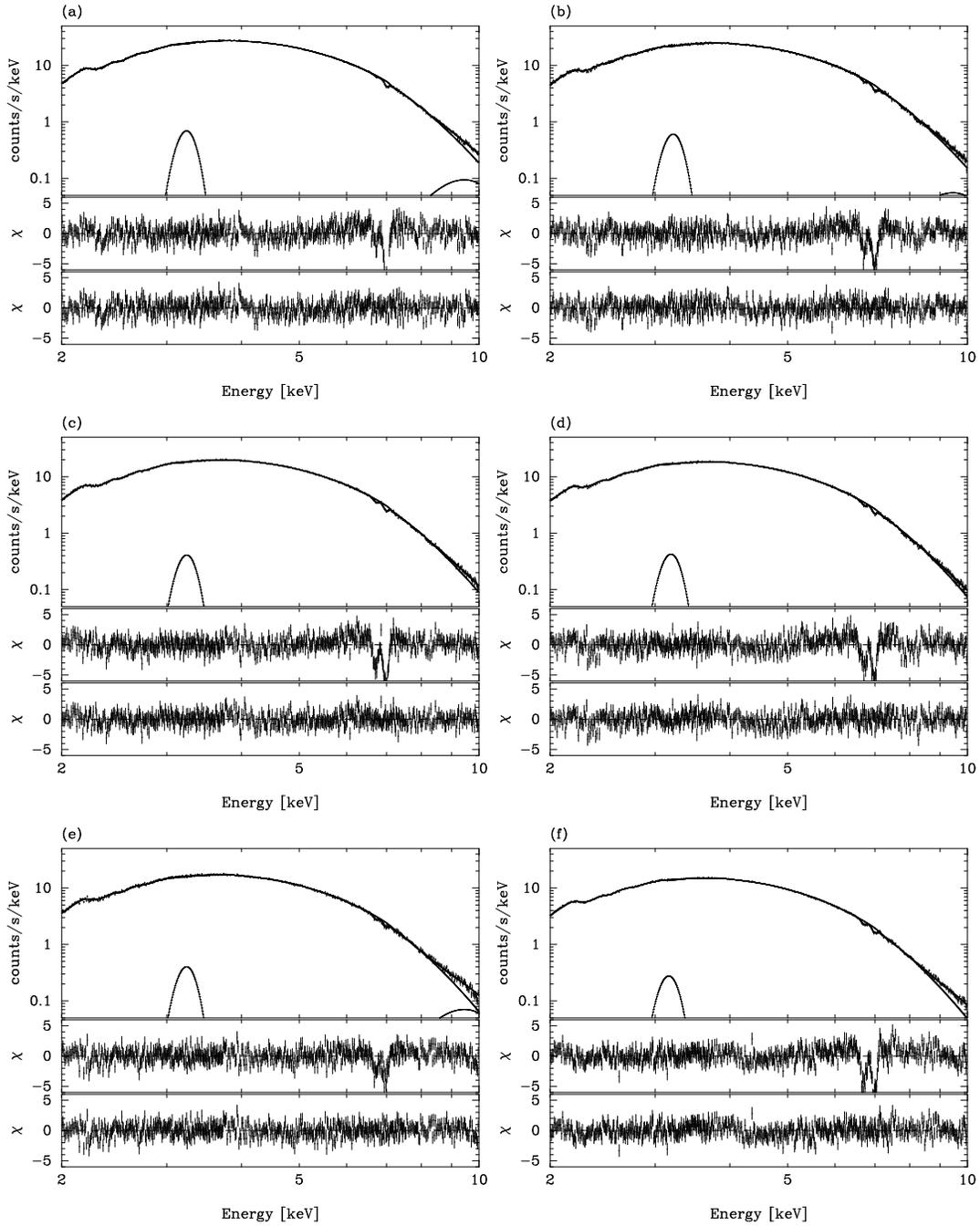

  \begin{center}

%
\FigureFile(70mm,60mm){figure3a.ps}
\FigureFile(70mm,60mm){figure3b.ps}
\FigureFile(70mm,60mm){figure3c.ps}
\FigureFile(70mm,60mm){figure3d.ps}
\FigureFile(70mm,60mm){figure3e.ps}
\FigureFile(70mm,60mm){figure3f.ps}
  \end{center}
  \caption{XIS023 spectra for the first to the last observations in panel (a) to (f) in this order.
  The observed data and the best-fit {\sc diskbb} model with negative gaussians are shown in the top
panels.
  The middle panels show residuals between the data and the model without 
  negative gaussians.  The bottom panels show the residuals between the data and the best-fit model.
Note that the burst option was not used in the
5th and 6th observations (Table~\ref{tab:obs}), hence the pileup effect is more obvious in 
panels (e) and (f) in spite of the gradual decrease of the source flux.
An edge-like structure may be indicated at $\sim4$~keV of which residual level is within 1\%. 
This structure is understood as 
a coupling of calcium K edge at 4.05~keV of interstellar absorption and the uncertainty of 
current response matrix.
}
  \label{fig:spec_xis2-10}
\end{figure*}

Figure~\ref{fig:spec_xis2-10} shows the 2--10~keV XIS raw spectra.  To
reproduce the continuum spectra, we use
the multi-color disk model ({\sc diskbb} in 
{\sc xspec}\footnote{http://heasarc.gsfc.nasa.gov/docs/xanadu/}, 
\cite{mitsuda84, max86}) 
modified by interstellar absorption 
({\sc wabs} model; \cite{mm83}), 
following the canonical modeling for the
high/soft state black hole binaries.  
Here, the {\sc diskbb} model approximates emission from the standard accretion disk, and 
is known to successfully reproduce the dominant soft spectra of the high/soft state black holes.
As shown in figure~\ref{fig:rawspec_xis-pin}b, the contribution of the power-law
tail is negligible in the XIS energy band, so is not included in the spectral fitting.
The residual effects of pileup are modeled
 by a broad gaussian at 10~keV (see \S~2.2),
and the current uncertainties in the
instrument response at the gold M-II edge by another gaussian at 3.2~keV (see \S~2.3).

We start with the first observation, which showed the brightest spectrum.
The continuum is well reproduced by the {\sc diskbb} model with a disk
temperature of $kT_{\rm in}=1.39$~keV. 
This is amongst the highest disk temperatures 
observed with RXTE
from this source in previous high/soft states (e.g., \cite{abe05}).  
A disk bolometric luminosity of $L_{\rm disk} = 2.8\times10^{38} \cdot a ~{\rm erg~s^{-1}}$  and 
an apparent disk inner radius of $r_{\rm in} = 24\cdot\sqrt{a}$~km are 
obtained,   
where $a$ is defined as 
$(D/10~{\rm kpc})^2/(\cos i/\cos 70^\circ)$.  The absorbing column density, $N_{\rm H}$, is
estimated to be $8.15\times 10^{22}~{\rm cm^{-2}}$.  This is 
close to that seen with previous observations by ASCA and RXTE (e.g.,
\cite{abe05}).

The fit with this simple model is however far from acceptable 
($\chi^2/dof=1465.0/574$) because of the complex structure in the iron K band.
Residuals between the data and the best-fit continuum model are 
shown in the middle panel of figure~\ref{fig:spec_xis2-10}a.
The most obvious features are two narrow dips at 7.0~keV and
6.8~keV.  We therefore add two negative gaussians to the model.  The fit is
dramatically improved with $\chi^2/dof=714.0/570$.  The line center
energies are estimated to be $7.001^{+0.006}_{-0.005}$~keV and $6.73\pm0.02$~keV, 
consistent with the H-like and He-like iron K$\alpha$ lines at 6.966~keV
and 6.697~keV~\footnote{
Weighted averages of K$\alpha_1$ and K$\alpha_2$
by referring to the National Institute of Standards and Technology
(NIST; see http://aeldata.phy.nist.gov/archive/data.html )
and \citet{drake88} for H-like and He-like K$\alpha$, respectively.};
the lines are possibly blue-shifted with
$z\sim 5\times 10^{-3}$.
%
%
The equivalent widths are 30~eV and 7~eV, for the H-like and
He-like lines, respectively.  
The lines are narrow with an upper limit to $\sigma$ of 40~eV and
50~eV. These values are consistent with what we expect for 
narrow lines given the current response matrix, 
which does not include the degradation of the energy resolution (\S~2.3). 
We thus fix the value of $\sigma$ at 10~eV to represent intrinsically narrow lines in this subsection. 

In addition to these obvious line structures, the residuals reveal weaker   
absorption structures at 7.8~keV and 8.2~keV. These energies are
very similar to the rest frame energies of He-like and H-like iron K$\beta$ lines, namely
7.88~keV and 8.25~keV. However, they are also close in energy to the He-like
and H-like nickel K$\alpha$ lines at 7.80~keV and 8.09~keV, respectively, 
which can have comparable equivalent width to that of the
iron K$\beta$ lines (Kotani et al 2000). The data can marginally
constrain two weak lines at these energies, but not four. Hence we
include two additional negative gaussians, again with the width $\sigma$
fixed at 10~eV, and allow the energies to vary freely. 
This decreases $\chi^2/dof$ to $686.4/566$,
indicating that the two negative gaussians are statistically significant at
the 99.98\% confidence level using an $F$-test (where
$F(\nu,\Delta\nu)\equiv (\Delta \chi^2/\Delta \nu)/\chi^2_\nu = 5.65$).
The best-fit parameters are summarized in Table~\ref{tab:gaussfit}.
Residuals between the data and the best-fit model are shown in the
bottom panel in figure~\ref{fig:spec_xis2-10}a. 
The inferred energies are almost consistent with 
H-like and He-like iron K$\beta$, though they can be contaminated by nickel K$\alpha$.
%

\begin{figure}
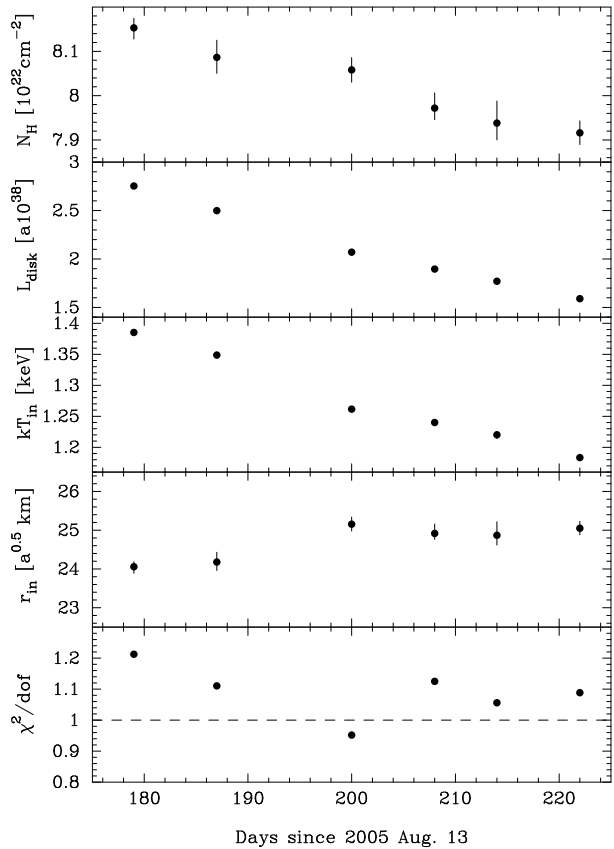

  \begin{center}
\FigureFile(80mm,40mm){figure4.ps}
  \end{center}
  \caption{Time histories of continuum parameters. Values of $N_{\rm H}$, $L_{\rm disk}$, $kT_{\rm in}$, $r_{\rm in}$~, and $\chi^2/{\rm d.o.f}$ are plotted.
  As described in the text, the decrease of $N_{\rm H}$ could be caused artificially by 
  the over-simplified {\sc diskbb} model. }
  \label{fig:disk}
\end{figure}

We performed the same spectral analysis on the other five observations.
The XIS spectra, the best-fit models, and the residuals are shown in figure~\ref{fig:spec_xis2-10}b--f.
The best-fit continuum and absorption parameters are summarized in
Table~\ref{tab:gaussfit}, and 
the time histories of
the continuum parameters are given in figure~\ref{fig:disk}.  Through the six
observations over two months, the value of $r_{\rm in}$ remained
approximately constant at $25\cdot\sqrt{a}$~km, while the disk
temperature gradually decreased to 1.18~keV, corresponding to a
factor 1.75 decrease in the disk bolometric luminosity to $L_{\rm disk}
=1.6\times10^{38}\cdot a~{\rm erg~s^{-1}}$.  The constancy of
$r_{\rm in}$ is another characteristic feature of the high/soft
state of black hole binaries (e.g., \cite{tanaka95} and references therein), and
confirms that the object was indeed in the high/soft state.
The correlated decrease in the derived value of $N_{\rm H}$ 
by $\sim$3\% could be an artifact of the over-simplified
 {\sc diskbb} model;
it assumes that the disk radiates locally as a blackbody, but distortions from
this are suggested both observationally (e.g., \cite{kubota04}; \cite{abe05}) and theoretically (\cite{watarai00}; \cite{davis05}) from
such luminous 
binaries.

\begin{figure}
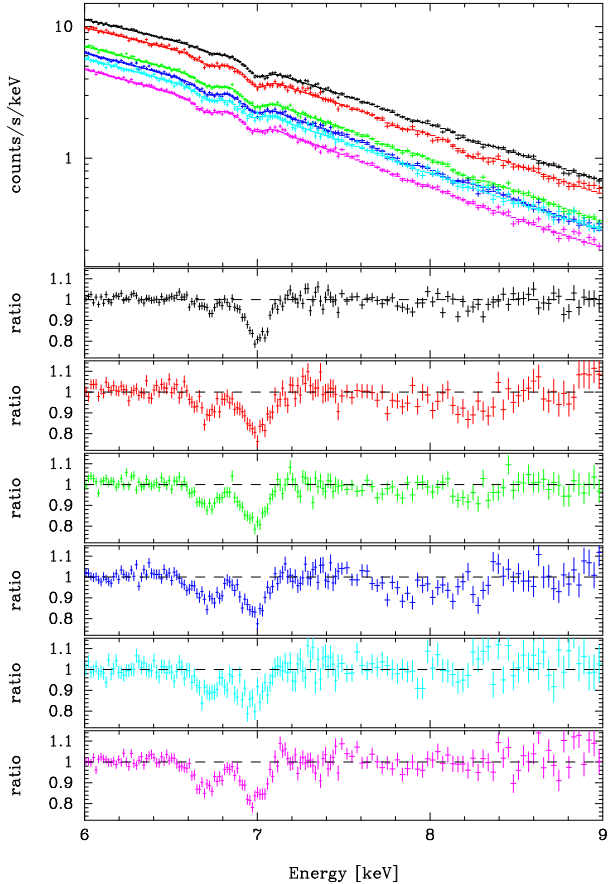

  \begin{center}
\FigureFile(80mm,40mm){figure5.ps}
  \end{center}
  \caption{The XIS023 6-9~keV spectra concentrating on the absorption line structure. Data and the best-fit models (Table~\ref{tab:gaussfit}) are shown in the top panel. Residuals between the data and the 
  best-fit model excluding the four negative gaussians are shown in the successive six panels.  
  Colors are the same as in figure~\ref{fig:rawspec_xis-pin}.
  }
  \label{fig:6-9}
\end{figure}

\begin{figure}
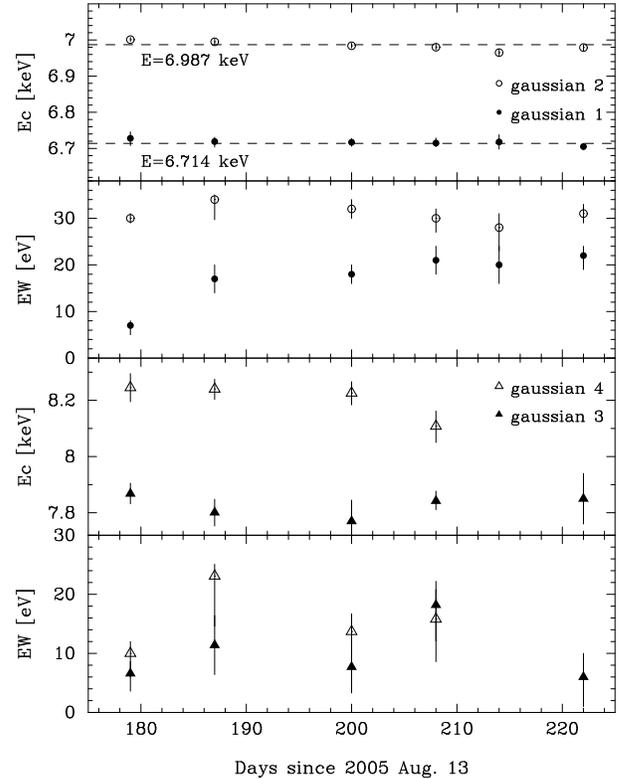

  \begin{center}
\FigureFile(80mm,40mm){figure6.ps}
  \end{center}
   \caption{Time histories of absorption line parameters based on the negative gaussian fit in the range of 6--10~keV.  The first two panels show line center energies and equivalent width of 
  gaussian 1 (filled circle) and gaussian 2 (open circle). The bottom two panels show
  those of gaussian 3 (filled triangle) and gaussian 4 (open triangle). On the top panel, average values are shown with dashed lines.}
  \label{fig:gauss}
\end{figure}

Irrespective of the continuum modeling, the two strong iron K$\alpha$
lines are clearly detected throughout the observations, 
whereas the weaker lines at higher energies are significantly detected in the
first four datasets ($F$-test significance values are given in
Table~\ref{tab:gaussfit}). 
Figure~\ref{fig:6-9} shows an enlargement of the 6--9~keV spectra, and 
residuals between the data and the continuum model.
The time histories of the individual
absorption-line equivalent widths and their energies are plotted in
figure~\ref{fig:gauss}.  While the equivalent width of the
H-like line was almost constant throughout the observations, that of the
He-like line increased by a factor of two between the first and second
observations, and then perhaps increased slightly more through the subsequent observations.  
These imply that the ionization state
of the absorbing gas decreased over time, 
as expected from the declining source luminosity.


The time averaged center energies of 
the two strong iron K$\alpha$ lines are estimated to be 
$6.987\pm0.005$~keV and $6.714\pm0.009$~keV, suggesting a blue shift 
velocity of $900\pm200~{\rm km~s^{-1}}$ and 
$700\pm400~{\rm km~s^{-1}}$, respectively.
If we take only the first observation, 
the blue shifts become more significant with $1500^{+300}_{-200}~{\rm km~s^{-1}}$ and
$1500\pm900~{\rm km~s^{-1}}$,
for the H-like and He-like irons, respectively. 
Here, the errors are statistical only.
Given the current uncertainty in the gain calibration of $\sim20$~eV or $\sim900~{\rm km~s^{-1}}$ 
(see \S~2.3), the blue shift cannot be measured with higher significance.
However,
a blue shift is physically expected, 
since 
the material responsible for the absorption lines
is certainly a wind from the disk (\S~1), 
and these values are consistent with the observed blue shift from other accreting systems like 
GRO~J$1655-40$ \citep{miller06} and GX~$13+1$ \citep{ueda04}.
In the next subsection, the absorption line parameters and their time
evolution are studied in detail.

\begin{table*}
  \caption{Best-fit parameters for the 2--10~keV XIS spectra}
  \label{tab:gaussfit}
  \begin{center}
    \begin{tabular}{lllllll}
  \hline \hline
 epoch &continuum  & gaussian 1& gaussian 2 & gaussian 3 & gaussian 4 & ${\chi^2/dof}$\\
	&$N_{\rm H}$~[$10^{22}~{\rm cm^{-2}}$]  &$E_{\rm c}$~[keV] &$E_{\rm c}$~[keV] &$E_{\rm c}$~[keV]&$E_{\rm c}$~[keV]&\\
	&$kT_{\rm in}$~[keV]&EW~[eV] &EW~[eV] &EW~[eV]&EW~[eV]&\\
\cline{5-6}
	&$r_{\rm in}$~[$\sqrt{a}$~km]&$(\sigma~[{\rm eV}])$~$^*$ & $(\sigma~[{\rm eV}])$~$^*$&\multicolumn{2}{c}{significance under $F$-test~$^\dagger$}&\\	
  \hline\hline
 1 &  $8.15^{+0.02}_{-0.03}$ &$6.73\pm0.02$ & $7.001^{+0.006}_{-0.005}$ &$7.87\pm0.04$ &$8.24\pm0.05$ & 686.3/566 \\
	  & $1.385^{+0.004}_{-0.003}$   &   $7^{+1}_{-2}$    &$30\pm1$     &$6.6_{-3}^{+2}(^{+0}_{-0.1})^\ddagger$  & $10.0_{-4}^{+2}(^{+0}_{-0.3})^\ddagger$     &\\
\cline{5-6}
& $24.1^{+0.1}_{-0.2}$  &$(<37)$&$(<49)$&$F(566,4)=5.69$&99.98\% &\\

\hline 
 2 &  $8.09\pm0.04$ & $6.72^{+0.01}_{-0.02}$ & $6.995^{+0.009}_{-0.008}$ &$7.80\pm0.05$ &$8.24^{+0.03}_{-0.04}$ &  628.6/566\\

 	  &$1.349^{+0.004}_{-0.005}$ &$17\pm3$    &$34_{-4}^{+1}$     &$11.4\pm5(^{+0}_{-0.1})$  & $23.1_{-8}^{+2}(^{+0}_{-0.5})$     &\\
\cline{5-6}
& $24.2^{+0.3}_{-0.2}$  &$(<46)$&$(13^{+36}_{-13})$&$F(566,4)=9.68$&$>99.999$\%& \\
 \hline
 3&  $8.06\pm0.03$ & $6.72\pm0.01$  &$6.984^{+ 0.007}_{-0.006}$   &$7.77\pm0.08 $ &$8.23\pm0.04$ &538.8/566  \\

 	  & $1.262\pm0.003$ & $18\pm2$ &$32\pm2$ &$7.74\pm4(^{+0}_{-0.04})$ &$13.7_{-5}^{+3}(^{+0}_{-0.2})$   &\\
\cline{5-6}
& $25.2^{+0.2}_{-0.2}$    &$(<37)$&$(18\pm18)$&$F(566,4)=6.89$&99.998\%& \\

\hline 
4&  $7.97\pm0.03$  & $6.72 \pm0.01$ &  $6.980^{+0.009}_{-0.007}$ &$7.84^{+0.04}_{-0.03}$ &$8.11^{+0.05}_{-0.06}$&  636.7/566\\

 	  &$1.240^{+0.003}_{-0.004}$  &  $21\pm3$ & $30_{-3}^{+2}$ &$18.2_{-6}^{+4}(^{+0}_{-0.1})$ &$15.8_{-7}^{+5}(^{+0}_{-0.2})$   &\\
\cline{5-6}
&$24.9^{+0.2}_{-0.2}$   &$(40^{+26}_{-21})$&$(<66)$&$F(566,4)=12.0$&$>99.999$\%& \\
\hline 
5&  $7.94\pm0.04$   & $6.72 \pm0.02$ &  $6.97\pm0.01$ &(7.8--7.9)$^\S$&(8.2--8.3)$^\S$&599.9/566 \\

 	  &$1.220^{+0.004}_{-0.005}$ &  $20\pm4$ & $28_{-5}^{+3}$ &$<13$&$<10$&\\
\cline{5-6}
&$24.9^{+0.3}_{-0.3}$  &$(<55)$&$(40^{+25}_{-40})$&$F(566,4)=0.34$&15\%& \\
 \hline
 6&  $7.92\pm0.03$   & $6.705\pm0.009$ &  $6.969^{+0.008}_{-0.006}$& $7.85\pm0.09$&(8.2--8.3)$^\S$&
 616.0/566  \\
 	  &$1.183\pm0.003$  &  $22_{-3}^{+2}$ & $31\pm2$ &$6^{+4}_{-5}$&$<9$  & \\
\cline{5-6}
&$25.1^{+0.2}_{-0.2}$   &$(22^{+25}_{-22})$&$(<23)$&$F(566,4)=1.05$&62\%& \\
\hline
   \multicolumn{7}{@{}l@{}}{\hbox to 0pt{\parbox{160mm}{\footnotesize
       Notes. Errors represents 90\% confidence limit of the statistical errors.  See \S 2.3 for systematic errors.
      \par\noindent
      \footnotemark[$*$] The best-fit values of $\sigma$, or the upper limit if the best-fit value is smaller than 10 eV.\par\noindent
      \footnotemark[$\dagger$] $F$-values are presented for the addition of gaussians 3 and 4.
            \par\noindent
      \footnotemark[$\ddagger$]The best fit normalization factors of the broad gaussian at 10~keV for the pileup are 
7.0,  4.1, 2.4, 1.9, 5.2, and $3.1\times 10^{-3}~{\rm photons~s^{-1}~cm^{-2}}$, in order of the 
first to the last observations. Uncertainty of the pileup effect to the equivalent width 
is considered by including/excluding the broad gaussian to the corresponding continuum.
            \par\noindent
      \footnotemark[$\S$] Line center energies are fixed at those of He-like and H-like iron K$\beta$ in the rest frame to obtain upper limits of the equivalent width, since they are not constrained by the fitting.
                }\hss}}
\end{tabular}
\end{center}
\end{table*}


\subsection{Reanalysis of the absorption lines with Voigt profile}

The simple gaussian analysis above shows significant He-like and H-like iron
K$\alpha$ absorption lines in the data, and suggests the presence of the weaker
absorption structures at 7.8--8.2~keV.
Since the parameters of the weaker high-energy lines are difficult to constrain, 
below we focus on the stronger two lines, and study them using 6.0--7.5~keV spectra.
The limited energy range means that we
do not need to include the (instrumental) broad gaussian at 10~keV or
the narrow gaussian at $3.2$~keV.  We model the continuum by {\sc
diskbb} as before, with free parameters $T_{\rm in}$ and $r_{\rm in}$, but fix
$N_{\rm H}$ to $8.0\times10^{22}~{\rm cm^{-2}}$, which is the average
value obtained from the 2--10~keV fits.
This does not affect the derived line center energies
or equivalent widths, 
since the differences of $N_{\rm H}$ from the fixed value, at most
$\sim1\times10^{21}~{\rm  cm^{-2}}$,  
produce a rather small change to the iron edge at 7.1~keV.


To conduct a detailed analysis of these absorption lines, we use a full
Voigt profile as implemented in the {\sc kabs} model. 
The {\sc kabs} model 
is constructed by \citet{ueda04} (see their \S 3.2.1) as a local model for {\sc xspec},
and has three fitting parameters for each specific ion;
the ion column density, $N_{\rm ion}$, 
the velocity dispersion, $b$, and the bulk outflow velocity characterized by a blue shift, $z$. 
The line shape is produced through a
convolution of a Lorentzian profile, describing natural and pressure
broadening 
with a gaussian describing Doppler motions (specified by the 
velocity dispersion including thermal and kinetic motion). 
Except for forbidden lines, each individual line of the K$\alpha$ multiplet
(K$\alpha_1$ and K$\alpha_2$) is included separately,
with the appropriate energy, oscillator strength, and Einstein coefficient.
This properly models the fine structure in the line
even though these details of the line profile cannot be resolved
by the CCD.
The values of chi-squared are the same as for the previous negative gaussian fits, but this
model gives the ion column $N_{\rm ion}$ 
for an assumed velocity dispersion, $b$, as opposed to simply an
equivalent width. We require the two lines to have the same $b$ and the same $z$, and neglect any
line emission which could fill in the absorption.

Although we cannot measure $b$ from line profiles, its upper and lower bounds may be obtained.
The upper limit is set at $b\sim2000~{\rm km~s^{-1}}$ by the upper limits on the line width, 
$\sigma\sim30$--40~eV (\S~3.2).
The lower limit to $b$
is given by the thermal velocity. At
such high ionization, the self-consistent temperature of the
photoionized material must be of order the Compton temperature, $T_{\rm C}$.
Considering the balance between Compton heating by the dominant disk 
illumination, and Compton, bremsstrahlung and ion coolings, 
$T_{\rm C}$ is roughly estimated as 
$kT_{\rm in}/3$ (i.e., 0.5~keV in this case).
The flux in the power-law tail has only a very small
effect on this as it is so much weaker than the disk (20--100~keV power-law flux is only 
$\sim1$\% of the disk bolometric flux), and does not contribute significantly on the 
compton heating (e.g., \cite{rey00}).
This temperature implies thermal (random kinetic) velocties as
 $b=\sqrt{2 kT_{\rm C}/ m_{\rm atom}}$, and thus $b=42~{\rm km~s^{-1}}$
can be obtained as a secure lower limit.

Table~\ref{tab:kabs} gives the derived $N_{\rm ion}$ for these
two limiting velocities, together with 
velocities of $b=100$, 200, and $500~{\rm km~s^{-1}}$. 
As expected from the
curve-of-growth analysis, the observed equivalent width can be produced by a 
lower value of $N_{\rm ion}$ if a higher value of $b$ is assumed. 
In Table~\ref{tab:kabs}, we can put further constraints on $b$ by using the observed
equivalent widths of weaker lines at 7.8~keV and 8.2~keV
(gaussian 3 and 4 in Table\ref{tab:gaussfit}).
The table includes the model predictions for the
equivalent width of the iron K$\beta$ absorption lines.
There, models with ``1" in the last column (``Note") are inconsistent, 
because the predicted iron K$\beta$ line equivalent widths exceeds the actually observed values, 
leaving no room for the nickel K$\alpha$ lines to contribute.
By scanning $b$ with finer steps,
we found that $b$ of $\sim60$--$200~{\rm km~s^{-1}}$ significantly 
over predicts the K$\beta$ equivalent width, and is hence ruled out.

The top three panels of figure~\ref{fig:kabs}  show the time histories of the best-fit {\sc kabs} parameters,
assuming $b=500~{\rm km~s^{-1}}$
which is based on the marginally resolved line width 
in the Chandra HETGS data of GX~$13+1$ \citep{ueda04}.
The column density of the H-like iron, $N_{\rm Fe~XXVI}$, is almost constant at 
$\sim1\times10^{18}~{\rm cm^{-2}}$ while that of 
the He-like iron, $N_{\rm Fe~XXV}$, increases significantly from the first to the second observation.
This is consistent with  the behavior of the equivalent width estimated using the negative
gaussian fit (Table~\ref{tab:gaussfit}).
The weighted average blue shift is about $z=(3.3\pm 0.7)\times 10^{-3}$, corresponding to 
an outflow velocity of 
$1000\pm200~{\rm km~s^{-1}}$.
Again, the first observation shows $z=(5.0^{+0.8}_{-0.6})\times10^{-3}$ or 
$v=1500^{+240}_{-180}~{\rm km~s^{-1}}$ (see Table~\ref{tab:kabs}).
They are consistent with the results obtained with the negative gaussian fits in \S~3.2.

\begin{table*}
  \caption{Best-fit parameters of the 6--7.5~keV {\sc kabs} model}
  \label{tab:kabs}
    \begin{tabular}{ccccccccccc}
  \hline\hline
  epoch & \multicolumn{4}{c}{The best-fit {\sc kabs} parameters}&\multicolumn{2}{c}{predicted K$\beta$ EW$^*$}&$N_{\rm ion}$ ratio$^\dagger$&\multicolumn{2}{c}{plasma parameters$^\ddagger$}&Note$^\S$\\
\cline{2-5}
\cline{9-10}
  & $N_{\rm Fe~XXV}$ &$N_{\rm Fe~XXVI}$ & $z^\|$ &$\chi^2$ & Fe XXV & Fe XXVI&He/H&$\xi$& $N_{\rm tot}$&\\
& $10^{18}~{\rm cm^{-2}}$ &$10^{18}~{\rm cm^{-2}}$ &  $10^{-3}$&$(96)^\#$  & eV& eV&$\times 10^{-2}$&$10^4$& $10^{23}~{\rm cm^{-2}}$&\\
\hline\hline
\multicolumn{11}{l}{$b=42~{\rm km~s^{-1}}$~~ ($kT_{\rm kin}=0.5$~keV)}\\
\hline
1 & $0.6^{+0.4}_{-0.3}$ & $34^{+ 6}_{-5}$ &$5.2^{+0.7}_{-0.7}$& 116.1&       4.1 &12.5     & $1.7^{+1.6}_{-1.0}$  &$19^{+15}_{-7}$ & $130^{+140}_{-60}$&2\\ 
2 & $4.9^{+2.2}_{-2.0}$ & $40^{+10}_{-9}$ &$4.2^{+1.0}_{-1.1}$& 97.9&         9.3    &11.3      & $12^{+10}_{-6}$  &  $5^{+4}_{-2}$ & $39^{+39}_{-16}$&2\\
3 & $4.4^{+1.6}_{-1.2}$ & $33^{+ 7}_{-6}$ &$2.8^{+1.0}_{-0.8}$& 98.8&      7.9    &12.6    & $14^{+9}_{-5}$  & $6^{+3}_{-2}$ & $29^{+18}_{-10}$&2\\ 
4 & $6.7^{+2.3}_{-2.0}$ & $31^{+ 7}_{-6}$ &$2.4^{+1.3}_{-0.7}$& 102.4& 10.1  &11.8       & $22^{+15}_{-9}$   & $5^{+2}_{-1} $& $22^{+12}_{-6}$&2\\ 
5 & $5.7^{+3.5}_{-2.3}$ & $27^{+11}_{-8}$ &$1.5\pm1.5$& 85.7&    8.1  & 10.3     &
 $21^{+28}_{-12}$  & $5^{+5}_{-2} $& $1.9^{+2.1}_{-0.7}$&2\\ 
6 & $7.7^{+2.5}_{-1.8}$ & $33^{+ 4}_{-6}$ &$1.9^{+1.0}_{-0.8}$& 108.2&    10.3  &12.0       & $24^{+15}_{-7}$   & $6\pm2$& $22^{+6}_{-5}$&1,2\\
 \hline \hline
 \multicolumn{11}{l}{$b=100~{\rm km~s^{-1}}$ ($kT_{\rm kin}=2.9$~keV)}\\
\hline
1 & $0.23^{+0.14}_{-0.09}$ & $24^{+ 6}_{-4}$ &$5.3^{+0.7}_{-0.7}$& 115.9&   3.1   & 17.1      & $1.0^{+1.0}_{-0.5}$  &$29^{+19}_{-11}$ &$140^{+130}_{-70}$ &1, 2\\ 
2 & $2.03^{+1.30}_{-0.80}$ & $31^{+10}_{-9}$ &$4.4^{+1.0}_{-1.1}$& 98.0&   11.3    & 17.6      & $7^{+9}_{-4}$  &$8^{+6}_{-4}$ & $45^{+58}_{-25}$&2 \\ 
3 & $1.98^{+0.97}_{-0.78}$ & $23^{+ 7}_{-6}$ &$3.0^{+0.9}_{-0.8}$& 99.2&     11.1  &   17.1    & 
$9^{+9}_{-5}$  & $9^{+6}_{-3}$& $26^{+30}_{-12}$&2\\ 
4 & $3.12^{+1.46}_{-1.07}$ & $21^{+ 7}_{-6}$ &$2.7^{+1.2}_{-0.8}$& 103.5&    12.9  &   16.8 & 
$15^{+16}_{-8}$  & $6^{+4}_{-3}$&$17^{+17}_{-8}$ &2\\ 
5 & $2.51^{+2.17}_{-1.17}$ & $17^{+10}_{-8}$ &$1.8^{+1.6}_{-1.6}$& 87.5&    11.9   &   16.4& 
$15^{+37}_{-10}$  & $7^{+8}_{-4}$&$14^{+27}_{-8}$ &1, 2\\ 
6 & $3.99^{+1.62}_{-1.21}$ & $22^{+ 7}_{-6}$ &$2.3^{+0.8}_{-0.9}$& 107.2&   13.9   &   17.0    &
 $18^{+16}_{-8}$  & $7^{+4}_{-2}$& $16^{+12}_{-6}$&1, 2\\ 
\hline \hline
   \multicolumn{11}{l}{$b=200~{\rm km~s^{-1}}$ ($kT_{\rm kin}=12$~keV)}\\
\hline
1 & $0.14^{+0.05}_{-0.04}$ & $3.6^{+1.2}_{-0.9}$ &$5.4^{+0.7}_{-0.7}$& 115.6&   2.2       &  15.6     & 
$4^{+3}_{-2}$  &$11^{+6}_{-4}$ &$8^{+8}_{-4}$ &1\\ 
2 & $0.66^{+0.34}_{-0.22}$ & $5.0^{+3.2}_{-1.8}$ &$4.3^{+1.0}_{-1.1}$& 99.8&  8.1 & 17.8    & 
$13^{+18}_{-8}$  &$5^{+4}_{-2}$ &$5^{+9}_{-3}$ &\\ 
3 & $0.65^{+0.13}_{-0.18}$ & $3.4^{+1.6}_{-1.0}$ &$3.0^{+0.9}_{-0.8}$& 98.9&  8.0 &  15.2   & 
$19^{+13}_{-10}$  &$5^{+3}_{-2}$ & $3^{+3}_{-1}$&\\ 
4 & $0.92^{+0.35}_{-0.26}$ & $3.0^{+1.4}_{-0.9}$ &$2.6^{+1.2}_{-0.7}$& 103.8&      9.9    &  14.3     & 
$30^{+29}_{-15}$  & $4^{+3}_{-1}$&$2.0^{+1.7}_{-0.7}$ &\\ 
5 & $0.77^{+0.49}_{-0.31}$ & $2.3^{+1.9}_{-0.9}$ &$1.5^{+1.7}_{-1.5}$& 87.2&  9.1 & 12.3   & 
$33^{+56}_{-22}$  & $4^{+5}_{-2}$& $1.5^{+2.5}_{-0.6}$&1\\ 
6 & $1.16^{+0.40}_{-0.30}$ & $3.4^{+1.5}_{-1.0}$ &$2.3^{+0.8}_{-0.8}$& 107.4&  11.6   &  15.0     &
 $35^{+30}_{-17}$  &$4^{+3}_{-2}$ &$2.2^{+1.5}_{-0.6}$ &1\\ 
\hline\hline
 \multicolumn{11}{l}{$b=500~{\rm km~s^{-1}}$~~($kT_{\rm kin}=72$~keV)}\\
\hline
1 & $0.11^{+0.03}_{-0.03}$ & $1.14^{+0.13}_{-0.12}$ &$5.0^{+0.8}_{-0.6}$& 115.6&1.9&8.5&
$10^{+4}_{-3}$& $5.6^{+1.9}_{-1.2}$& $1.3^{+0.6}_{-0.3}$&\\
2 & $0.29^{+0.08}_{-0.06}$ & $1.27^{+0.24}_{-0.20}$ &$4.0^{+1.0}_{-1.1}$& 99.8&4.7 &9.7&
$23^{+12}_{-8}$& $3.3^{+1.1}_{-0.9} $& $0.9^{+0.4}_{-0.2}$&\\
3 & $0.30^{+0.06}_{-0.03}$ & $1.11^{+0.17}_{-0.15}$ &$2.6^{+1.1}_{-0.8}$& 98.7&4.9&8.3&
$27^{+10}_{-6}$ &$ 3.8^{+0.8}_{-0.7}$& $0.8\pm0.2$&\\
4 & $0.36^{+0.08}_{-0.06}$ & $1.05^{+0.18}_{-0.15}$ &$2.2^{+1.3}_{-0.7}$& 103.7&5.4 &8.0&
$34^{+14}_{-10}$& $3.5^{+1.0}_{-0.8}$& $0.7^{+0.2}_{-0.1}$ &\\
5 & $0.34^{+0.10}_{-0.09}$ & $0.95^{+0.23}_{-0.20}$ &$1.1^{+1.6}_{-1.5}$& 86.3&5.4 &7.7&
$35^{+23}_{-15}$& $3.6^{+1.6}_{-1.1}$& $0.6^{+0.2}_{-0.1}$&\\
6 & $0.41^{+0.08}_{-0.07}$ & $1.10^{+0.18}_{-0.15}$ &$1.6^{+1.0}_{-0.3}$& 108.3&6.5 &8.5&
$37^{+14}_{-10}$& $4.0^{+1.1}_{-0.8}$&$0.7^{+0.2}_{-0.1}$&\\
\hline\hline
 \multicolumn{11}{l}{$b=2000~{\rm km~s^{-1}}$($kT_{\rm kin}=1200$~keV)}\\
\hline
1 & $0.11^{+0.01}_{-0.03}$ & $0.89^{+0.06}_{-0.06}$ &$5.0^{+0.7}_{-0.6}$& 123.7&      1.9 &7.2    & 
$12^{+2}_{-4}$  & $4.7^{+1.4}_{-0.5}$& $0.89^{+0.3}_{-0.1}$&\\ 
2 & $0.26^{+0.05}_{-0.04}$ & $0.97^{+0.13}_{-0.10}$ &$4.0^{+1.1}_{-1.1}$& 95.0&   4.1     & 8.5      & 
$27^{+9}_{-7}$  & $3.0^{+0.7}_{-0.5}$& $0.7^{+0.2}_{-0.1}$&\\ 
3 & $0.26^{+0.04}_{-0.04}$ & $0.87^{+0.09}_{-0.09}$ &$2.5^{+1.0}_{-0.8}$& 101.1&   4.5    & 7.1      & 
$30^{+8}_{-7}$  & $3.6^{+0.7}_{-0.6} $& $0.6\pm0.1$&\\ 
4 & $0.29^{+0.05}_{-0.04}$ & $0.84^{+0.06}_{-0.08}$ &$2.2^{+1.1}_{-0.9}$& 100.2&  4.9     &     7.5  & 
$35^{+10}_{-6}$  &$3.5^{+0.6}_{-0.6} $& $0.53\pm0.06$&\\ 
5 & $0.28^{+0.07}_{-0.05}$ & $0.80^{+0.13}_{-0.14}$ &$1.1^{+0.9}_{-1.6}$& 82.6&   4.4     &6.9       & $35^{+18}_{-11}$  & $3.7^{+1.0}_{-0.9}$& $0.5\pm0.1$&\\ 
6 & $0.32^{+0.04}_{-0.03}$ & $0.87^{+0.08}_{-0.08}$ &$1.6^{+1.1}_{-0.8}$& 110.4&   5.1    &6.9       & $37^{+9}_{-7}$  & $4.0^{+0.6}_{-0.6}$& $0.55\pm0.06$&\\ 
\hline
   \multicolumn{11}{@{}l@{}}{\hbox to 0pt{\parbox{160mm}{\footnotesize
       Notes. Errors represents 90\% confidence limit of statistical errors.  See \S~2.3 for systematic errors.
      \footnotemark[$*$] Predicted equivalent width of He-like and H-like iron K$\beta$ under the best-fit {\sc kabs} model.
       \footnotemark[$\dagger$] Ratio of $N_{\rm Fe~XXV}$ to $N_{\rm Fe~XXVI}$ from the best-fit values.
       \footnotemark[$\ddagger$] $\xi$-parameters and total column density are calculated based on equations (1) and (2) derived in \S4.
      \footnotemark[$\S$] The models are rejected based on overestimation of K$\beta$ equivalent width (denoted as 1) or large $N_{\rm tot}$ as $>1.6\times 10^{24}~{\rm cm^{-2}}$ (denoted as 2).
       \footnotemark[$\|$] Plus is defined as blue shift.
       \footnotemark[$\#$] Degrees of freedom.

                }\hss}}
 \end{tabular}
  
\end{table*}

\begin{figure}
  \begin{center}
   \FigureFile(80mm,40mm){figure7.ps}
  \end{center}
  \caption{Time histories of absorption line parameters based on the Voigt profile fitting for 
  $b=500~{\rm km~s^{-1}}$. 
The top panel shows column density of H-like (open circle) and He-like iron (filled circle) 
in unit of $10^{18}~{\rm cm^{-2}}$, 
the second panel shows $N_{\rm Fe~XXV}/N_{\rm Fe~XXVI}$, 
and the third panel shows the line blue shift, $z$, in unit of $10^{-3}$. 
Systematic uncertainty of $\sim20$~eV for the absolute energy is interpreted as 
the uncertainty of $z$ as $\Delta z\sim 3\times10^{-3}$ at 7~keV, which is
also shown in the third panel.
Based on the ratio, $\xi$-parameter and the total hydrogen column $N_{\rm tot}$ are calculated and are 
shown in the next two panels in units of $10^4$ and $10^{23}~{\rm cm^{-2}}$, respectively. 
The best estimation of  $\xi$ and $N_{\rm tot}$ for $b=200~{\rm km~s^{-1}}$(dashed line)  
and $2000~{\rm km~s^{-1}}$ (dash-dot line) are
also shown in the same panels. }
  \label{fig:kabs}
\end{figure}

\section{Physical parameters of the absorber}

We observed very significant absorption lines from He--like and H--like
iron K$\alpha$ in the disk dominated X-ray spectrum of 4U~$1630-472$
during its decline from outburst.
The inferred blue shift of the absorption lines supports 
the presence of an outflow.
This source thus joins
the growing number of Galactic binaries with such absorption features,
suggesting that winds are a generic feature of bright accretion
disks observed at high inclination angles.
Through the analyses in \S~3, 
we found that 
iron is always
predominantly H--like, requiring that the ionization parameter,
$\xi=L/nr^2$ (where 
$n$ and $r$ are the number density and the distance of the absorbing 
material from the illuminating source), be very high.
In this section, we use the {\sc xstar} photoionization
code\footnote{http://heasarc.gsfc.nasa.gov/docs/software/xstar/xstar.html}
(version 2.1kn5, \cite{kallman01}) to calculate the ionization balance
of the  line-producing material, to convert the ion column densities into physical parameters
including total column density, number density, and location, for each observation.

\begin{figure*}
\begin{center}
\FigureFile(80mm,60mm){figure8a.ps}
\FigureFile(80mm,60mm){figure8b.ps}
 \end{center}
  \caption{Relative ionization population, $f_{\rm ion}$, of iron (a) and 
  the ratio of He-like to H-like iron (b), shown as a function of the ionization parameter $\xi$.
   Photoionization-recombination 
  equilibrium is assumed in an X-ray irradiated matter with a small optical depth.
  The irradiating X-ray spectral shape is assumed to be a {\sc diskbb} of $kT_{\rm in}=1.4$~keV (solid line) and $kT_{\rm in}=1.2$~keV (dashed line).
  }
  \label{fig:fepop}
\end{figure*}

\subsection{First observation}

We derive the physical parameters of the absorber 
using the first  observation as a specific example to illustrate the method.
We calculate the relative ion populations under illumination by a 
{\sc diskbb} spectrum, and use this to predict a theoretical ratio of 
$N_{\rm Fe~XXV}$ to $N_{\rm Fe~XXVI}$.
 The observed ratio from the data then enables us to 
estimate the $\xi$-parameter, and so convert
the observed ion column  densities of H-like and He-like irons into a 
total column density, $N_{\rm tot}$.
  
We use the illuminating spectrum derived from the Suzaku data, i.e., a
{\sc diskbb} spectrum of $kT_{\rm in}=1.4$~keV, to calculate the
self-consistent ion populations of iron using {\sc xstar}. 
The solid lines in figure~\ref{fig:fepop}a shows the calculated relative
fractions of each ionization stage of iron, $f_{\rm ion}$, against $\xi$, while
figure~\ref{fig:fepop}b shows $N_{\rm Fe~XXV}/N_{\rm Fe~XXVI}$.  
The ratio can be well approximated by a power law, as
$N_{\rm Fe~XXV}/N_{\rm Fe~ XXVI}\approx 4\times 10^5\cdot \xi^{-1.4}$.
Similarly, we obtain 
the fraction of H-like iron as
$f_{\rm XXVI}\approx 1.2\times10^4 \cdot \xi^{-0.98}$
at high values of $\xi$ (roughly $\xi>3\times10^{4}$).
Since a value of $N_{\rm tot}$ is obtained as $N_{\rm tot} \cdot
A_{\rm Fe}\cdot f_{\rm XXVI}=N_{\rm Fe~ XXVI}$, where $A_{\rm
Fe}=3.3\times 10^{-5}$ is the solar iron abundance, 
it can be estimated as 
$N_{\rm tot}=N_{\rm Fe~XXVI}/(A_{\rm Fe}\cdot f_{\rm XXVI})$.
By using these relations, we derive 
$\xi$ and $N_{\rm tot}$ for the range of $b$ given in Table~\ref{tab:kabs}.
The derived values of $\xi$, $N_{\rm tot}$ and $N_{\rm XXV}/N_{\rm XXVI}$ 
are also summarized in the same table.
In general, all these quantities are anti-correlated with $b$, and
at low velocity dispersions of $b=42$ and $100~{\rm km~s^{-1}}$, 
$N_{\rm tot}$ exceeds $1.6\times 10^{24}~{\rm cm^{-2}}$ 
i.e., is optically thick to electron scattering. 
This case is indicated as ``2" in the last column (``Note") of the table.
Thus, the case of $b\le100~{\rm km~s^{-1}}$ is ruled out by both 
the K$\beta$ absorption structures and the condition of optically thin scattering.
We can thus conclude that there must be
higher velocities present than expected simply from the temperature of the material, pointing to the importance of either a structured velocity field as expected in an
accelerating outflow, and/or turbulence.

\subsection{Time evolution of the absorber}

We repeat the above analysis for all the datasets. 
Since the value of $kT_{\rm in}$ changes as the source
    decline,  we calculate {\sc xstar} models $T_{\rm in}$ spanning the observed
    range of 1.2, 1.3, and 1.4~keV, 
and then use this to interpolate 
the ion populations for each observed temperature. The dashed lines
in figure~\ref{fig:fepop} show the results with $kT_{\rm in}=1.2$~keV, in comparison with
the solid lines which assumes $kT_{\rm in}=1.4$~keV. 
We approximated the behavior of the ratio of He-like to H-like ion columns 
and $f_{\rm XXVI}$ by following two formulae, 
\begin{equation}
N_{\rm Fe~XXV}/N_{\rm Fe~XXVI}=4\times 10^5\cdot 
(kT_{\rm in}/1.4)^{-5.57}\cdot \xi^{-1.4}~~,
\end{equation}
\begin{equation}
\log (f_{\rm XXVI})=-2+1.69\cdot \exp \left( \frac{-(\log (\xi/\xi_0))^2}{2} \right)~~,
\end{equation}
where $f_{\rm XXVI}$ peaks at $\log (\xi_0)=4.13\cdot(kT_{\rm
in}/1.4)^{-0.53}$.  Using these formulae, we converted the observed values of
$N_{\rm XXV}/N_{\rm XXVI}$ and $N_{\rm XXVI}$ into $N_{\rm tot}$ and $\xi$,
as given in Table~\ref{tab:kabs}.  For all the data sets, $b=42~{\rm km~s^{-1}}$ 
($kT_{\rm kin}=0.5$~keV)
and $b=100~{\rm km~s^{-1}}$ are again ruled out because $N_{\rm tot}$
becomes optically thick, 
as well as generally over-predicting the iron K$\beta$ absorption (\S~3.3).


The inferred time histories of the values of $\xi$ and $N_{\rm tot}$
are shown in figure~\ref{fig:kabs} assuming
$b=500~{\rm km~s^{-1}}$, together with upper and lower bounds corresponding to 
 $b=200$ and $2000~{\rm km~s^{-1}}$.  Assuming that the velocity
dispersion stays constant at $b=500~{\rm km~s^{-1}}$ then there is a
marginal decrease in $N_{\rm tot}$ as the luminosity
decreases, from $\sim 1\times10^{23}~{\rm cm^{-2}}$ in the first
observation to $\sim 7\times 10^{22}~{\rm cm^{-2}}$ in the final
dataset, as well as a marginal decrease in $\xi$ from
$\sim6\times 10^4$ to $\sim 4\times 10^4$.

\subsection{Location of absorbers}
We can roughly estimate $n$ and 
$r$ of the absorbing plasma, by using the definitions of 
$N_{\rm tot}=\int_{R_0}^\infty n\cdot dr$ and $\xi=L/nr^2$ where $R_0$ is the launching radius of the
absorber.
In this section, we use 
$b=500~{\rm km~s^{-1}}$ to estimate $\xi$ and $N_{\rm tot}$, 
and choose to replace
luminosity with flux, so as to make the dependence on the poorly
constrained source distance explicit.
We parameterize this in terms of
$D_{10}$, denoting distance in units of 10~kpc, and note there is no
uncertainty in $\xi$ parameter 
associated with the inclination,
since the absorber is along the same line of sight as we are.

The description of $N_{\rm tot}$ can be simplified as $N_{\rm tot}=\bar{n}\Delta R$ with 
a mean number density $\bar{n}$ and a characteristic thickness of the absorber $\Delta R$.
Using $\xi=L/\bar{n}R^2$, 
the average distance of the absorber, $R$, is thus
estimated as $R\sim 3\times10^{10}\cdot (\Delta R/R)\cdot {D_{10}}^2$ 
and $\sim 4\times10^{10}\cdot (\Delta R/R)\cdot {D_{10}}^2~{\rm cm}$,  
for the first and the last observations, respectively.
The densities are also estimated as 
$\bar{n}\sim 5\times 10^{12}\cdot (\Delta R/R)^{-2}\cdot {D_{10}}^{-2}$ 
and $\sim 2 \times 10^{12}\cdot (\Delta R/R)^{-2}\cdot {D_{10}}^{-2}~{\rm atoms~cm^{-3}}$.
Assuming $\Delta R/R\sim 1$ by following the previous research 
(e.g., \cite{kotani00,ueda01}; 2004), 
the values of $R$ and $\bar{n}$ are
estimated to be on the order of $R\sim10^{10}\cdot {D_{10}}^2~{\rm cm}$ and 
$\bar{n}\sim 10^{12}\cdot {D_{10}}^{-2}~{\rm atoms~cm^{-3}}$. 
They depend on the plasma thickness $\Delta R/R$, and
the longer distance and higher density are realized on the 
the condition of smaller $\Delta R/R$.
However, since this structure is continuous, it seems unlikely to have too small thickness
(e.g., $\Delta R/R < 0.1$).
Thus the range of the distance can be securely estimated as $10^{9-10}\cdot {D_{10}}^2~{\rm cm}$.
This value is well consistent with those estimated from the other binaries which showed 
absorption line structures. 

%
%

Because of lack of any evidence for low ionization material, 
it is reasonable and convenient to assume a constant ionization state
throughout the matter.
We thus assume that $n$ drops as $r^{-2}$ to keep $\xi$ constant at different distances.
This is also physically realistic for a geometrically thick absorber,
since mass conservation requires this dependency
for a constant outflow velocity with constant opening angle.
In this case, $N_{\rm tot}$ and $\xi$
are described as 
$N_{\rm tot}=n_0R_0$ and $\xi=L/n_0{R_0}^2$,
respectively, where $n_0$ is the launching density at $R_0$.
The launching distance  is thus estimated as 
$R_0\sim (3$--$4) \times10^{10}\cdot {D_{10}}^2$~cm throughout the observations, 
with the associated density of $n_0\sim (5$--$2)\times 10^{12}\cdot D_{10}^{-2}$ cm$^{-3}$.
These values are in good agreement with the previous estimation of the mean distance $R$ as expected as averaging the density profile gives an effective width
of $\Delta R/R_0\sim0.7$. 
There are no significant changes over the six observations, but the data are consistent with a small
increase in $R_0$ and/or a small decrease in $n_0$ as the 
luminosity decreases. 
There are some uncertainties on these numbers due to the statistical
uncertainties in $N_{\rm tot}$ and $\xi$, but the larger source of uncertainty
is systematic from the assumed velocity dispersion of 
$b=500~{\rm km~s^{-1}}$. The lower 
limit on  $b$ gives an upper limit on 
$N_{\rm tot}$ and $\xi$, and thus 
results in smaller launching radius
with larger density.  Considering our range in
$b$ of 200--$2000~{\rm km~s^{-1}}$,
the allowed range of $R_0$ is estimated to be 
(0.2--4)$\times10^{10}\cdot {D_{10}}^2$~cm and (1--5)$\times10^{10}\cdot {D_{10}}^2$~cm for the first and the
last observations, respectively.

\section{The nature of the wind}

The estimated launch radii are much smaller than the distance traveled by the
absorbing material over a single 20~ks observation from the measured
blue shift of $\sim 1000~{\rm km~s^{-1}}$, $10^{12}\cdot D_{10}^2~{\rm cm}$.
Thus the persistence of the wind
over the two-month monitoring period, 
where the
source was always in the high/soft state, shows that the material is not associated with a 
one--off event such as a state transition.
Neither can it be connected to the jet since the 
radio emission
is strongly suppressed in the high/soft state (Fender et al. 1999). 
Instead it must be produced
by the disk in a continuous or quasi-continuous manner.

Here we use the estimated physical parameters of the absorber to
constrain the driving mechanism for this wind.
One of the candidate mechanisms is radiation pressure on electrons, which
becomes dynamically important
as $L$ approaches $L_{\rm Edd}$, reducing the effective gravity by a factor
$\sim (1-L/L_{\rm Edd})$. This can be made much more efficient if the
cross-section for interaction between the matter and radiation is
enhanced by line opacity. There are multiple line transitions in the UV
region of the spectrum, so UV emitting disks can drive a powerful wind
at luminosities far below Eddington. Such line driven disk winds are
seen in CV's (\cite{Pereyra00}) and are probably also responsible for
the broad absorption line (BAL) outflows seen in AGN (\cite{proga00}).
However, the disk temperature for black hole binaries is in the X-ray
regime, so line driving is probably unimportant here (Proga \& Kallman
2002). We do indeed see lines, but the momentum absorbed in these iron
transitions is very small, completely insufficient to drive the wind.

Another type of outflow from a disk is a thermally driven wind
(\cite{begelman83}). Here again the central illumination is important,
but the process is less direct.  The illumination heats the upper layers
of the disk to a temperature of order the Compton temperature, $ T_{\rm C}$. This will
expand due to the pressure gradient, and at large enough radii the
thermal energy driving the expansion is larger than the binding energy,
leading to a wind from the outer disk. Simple estimates of the launching
radius of this wind give $R=10^{10} \cdot (M/M_\odot)\cdot ( T_{\rm C}/10^8~{\rm K})^{-1}
\sim 10^{12}$ cm (\cite{begelman83}).
This is a much larger distance than inferred for
our wind. However, this is an overestimate, and a more careful
analysis shows that thermal winds are launched at a radius a factor 5--10
smaller than this (\cite{begelman83}; \cite{woods96}) .
Secondly, the {\em combination} of thermal driving
and radiation pressure from moderately sub-Eddington illumination can be much more effective at
producing a wind. The hydrodynamic calculation of Proga \& Kallman (2002)
shows that thermal driving with $L/L_{\rm Edd}=0.6$
gives a powerful wind from well
within the expected thermal driving radius.  Their simulation is not
exactly matched to the situation here, as they model a neutron star
rather than a black hole. Nonetheless, their wind is launched from a
distance of around $1000R_{\rm s}$, has a column between $10^{22-23}$ cm$^{-2}$
and ionization of $\xi\sim 10^5$ for inclinations larger than $70^\circ$.
This is very close to our inferred parameters for the wind seen here in 4U~$1630-472$.
The simulation even has an outflow velocity of $\sim 1000~{\rm km~s^{-1}}$ as currently
indicated by the data.

The last type of outflow is a magnetically driven wind. These are much
harder to quantitatively study as the magnetic field configuration is
not known, yet they are almost certainly present at some level as the
underlying angular momentum transport is known to be due to 
magnetic fields (see e.g. Balbus \& Hawley
2002).  Winds (and jets) are clearly present in magnetohydrodynamical
(MHD) simulations which include these magnetic stresses self
consistently. 
These generically show that the mass loss is stochastic, with large fluctuations
both spatially and temporally, but that the time averaged properties are well defined, so this
magnetic wind is quasi-continuous (e.g., \cite{hawley01, machida04}). 
These calculations are still in their infancy, especially for
describing the properties of a geometrically thin disk as appropriate here, so cannot yet be used
for quantatiative predictions of the wind in the high/soft state. Instead, an approximation to
the properties of the self-consistent magnetic wind from an accretion disk can be made by
imposing an external field geometry (Proga 2000; 2003).
The mass loss rates depend
on this field configuration, but in general 
these allow steady, powerful winds to be
launched from any radius.  This lack
of diagnostic power means that they can be reliably identified only when
all other potential mechanisms for a wind are ruled out.

While the physical parameters required for the absorbing material seen
in 4U~$1630-472$ are somewhat more extreme than produced by the simple
analytic estimates of a thermally driven wind, they seem to be
potentially within the range of composite thermal/radiation pressure
driven winds if $L/L_{\rm Edd}>0.1$. Unfortunately there are no firm
estimates for the distance and black hole mass for 4U~$1630-472$, and
thus the Eddington ratio ($L/L_{\rm Edd}\sim 0.2$--$0.3\cdot a$) is
hard to constrain.
Neither are there any directly applicable simulations
of these winds, and how their properties might change with $L/L_{\rm Edd}$
decreasing by a factor 1.7.
Hence there is no unambiguous requirement for
magnetic driving from these data. Nonetheless, this wind is clearly
very similar in properties (though with a slightly higher ionization
state) to that that seen in a similar high/soft state from another
black hole, GRO J1655-40, where \citet{miller06} inferred a very small
launching radius from the Chandra grating data requiring a magnetic
wind. Given the continuity of properties seen between the wind in
4U~$1630-472$ and GRO~J$1655-40$ then it seems likely that if the wind
in GRO~J$1655-40$ is magnetically driven, then this process is
important also in the 4U~$1630-472$ absorber examined here.


%

\vspace{8mm}
We are grateful to all the Suzaku team members. 
We also thank Tim Kallman for making the {\sc xstar} package publicly available.
A.K. is supported by a special postdoctoral researchers program in RIKEN\@.
T.K. is supported by the Japan-Russia Research Cooperative Program of
JSPS\@.
C.D. thanks ISAS
for hospitality during the period over which this work was carried out.
The present work is supported in part by Grant-in-Aid No.1703011 from Ministry of 
Education, Culture, Sports, Science and Technology of Japan. 



\end{document}